# Formal verification for robo-advisors: Irrelevant for subjective end-user trust, yet decisive for investment behavior?

Tausch, Alina[1][2]

Wischnewski, Magdalena[3]

Yalciner, Mustafa[3]

Neider, Daniel[3]

Corresponding author: Alina Tausch

[1] Ruhr University Bochum, Universitätsstraße 150, 44801 Bochum, Germany

[2] Witten/Herdecke University, Alfred-Herrhausen-Straße 50, 58455 Witten, Germany | alina.tausch@uni-wh.de

[3] Research Center Trustworthy Data Science and Security, TU Dortmund University, Joseph-von-Fraunhofer-Straße 25, Dortmund, Germany

**Funding:** This project received a seed funding by the Research Center Trustworthy Data Science and Security (RC Trust) in 2023, which is funded by the Ministry of Culture and Science of the State of North Rhine-Westphalia and is part of the Research Alliance Ruhr.

**Abstract**

This online-vignette study investigates the impact of certification and verification as measures for quality assurance of AI on trust and use of a robo-advisor. Confronting 520 Prolific participants with an imaginary situation where they were using an online banking service to invest (parts of) their inherited money, we formed 4 experimental groups. EG1 achved no further information of their robo-advisor, while EG2 was informed that their robo-advisor was certified by a reliable agency for unbiased processes, and EG3 was presented with a formally verified robo-advisor that was proven to consider their investment preferences. A control group was presented a remote certified human financial advisor. All groups had to decisde on how much of their 10.000€ they would give to their advisor to autonomously invest for them and report on trust and perceived dependability. A second manipulation happened afterwards, confronting participants with either a successful or failed investment. Overall, our results show that the level of quality assurance of the advisor had surprisingly near to no effect of any of our outcome variables, except for people's perception of their own mental model of the advisor. Descriptively, differences between investments show that seem to favor a verified advisor with a median investment of 65,000€ (vs. 50,000€). Success or failure information, though influences only partially by advisor quality, has been perceived as a more important clue for advisor trustworthiness, leading to substantially different trust and dependability ratings. The study shows the importance of thoroughly investigating not only trust, but also trusting behavior with objective measures. It also underlines the need for future research on formal verification, that might be the gold standard in proving AI mathematically, but seems not to take full effect as a cue for trustworthiness for end-users.

# Formal verification as a means to ensure reliability and reliance?

In computer science, mathematical proofs of the reliability of algorithms that are generated by using formal methods are an established measure for hardware, software and, lately, for AI systems (Albarghouthi, 2021). They allow formally verifying certain qualities of algorithms, thus formally ensuring reliability regarding those qualities. A prototypical example is the robustness of AI systems, which stipulates that small perturbations to the input of a neural network should not cause large changes in its output. Using automated mathematical proof techniques, one can verify (i.e., mathematically guarantee) that a neural network is robust (or produce inputs that violate this property).

From a psychological perspective, formal proofs should be the ultimately useful signals for trustworthiness. Still, we do not yet know if the application of mathematical proofs to verify AI algorithms actually leads to an increase in trust and reliance on the part of end-users.

Such mathematical proofs are mostly complex and intransparent and can only (partially) be understood by mathematical experts. As such, they are just as intransparent and incomprehensible as the AI system itself to non-experts. Still, those systems can and will increasingly be used by such non-experts as part of their work routines or in their daily lives, seeking advice and decision support or even delegating decision-making. This is increasingly the case, e.g., in finance, with robo-advisors becoming more interesting, especially for those with some financial literacy and for online financial activities (Isaia & Oggero, 2022). Those robo-advisors offer access to investment help at lower costs than traditional banking services involving humans (D'Acunto et al., 2019) and as such are aimed at private usage.

Combining research knowledge on fundamental mathematics with ideas on the utilization of AI and psychological insights on trust in interaction with technologies, we jointly answer the *question whether (incomprehensible, yet appropriate) mathematical proofs lead to a form of trust and reliance using a robo-advisor.* In more detail, we want to compare a robo-advisor that is proven in considering people's financial preferences to other robo-advisors being (1) certified by a reliable institution or (2) not providing information on process or outcome standards as well as to (3) the classical human financial advisor. Doing so, our aim is to understand how varyingly valuable cues for actual trustworthiness (see Realistic Accuracy Model by Funder, 1995) are building one's perceived trustworthiness and how they are reflected in reliance behavior.

## 1 Theoretical Background: Trust in AI-driven investment support

In their study investigating the impact of artificial intelligence (AI) in Financial Services, PricewaterhouseCoopers (2020) describe the aim of the sector to apply AI for changing processes as well as structures to foster efficiency or support the development towards customer experiences. One application towards the latter goal is the so-called robo-advisors. These systems are artificially intelligent financial advisors based on processing of data (trained on huge quantities of financial data) that offer advice regarding investments without humans involved (Bhatia et al., 2020; Hodge et al., 2021; Tao et al., 2021). The strength of robo-advisors lie in the inherent neutrality of their advice in comparison to individual humans (D'Acunto et al., 2019), as well as in their continuous availability (Cardillo & Chiappini, 2024). Still, research in this area, especially considering individual outcomes, is only growing since 2019 (D'Acunto et al., 2019) – and questions regarding the willingness of people to invest using such robo-advisors and the conditions under which they are trusted need to be thoroughly investigated.

## 1.1 Trust & Trustworthiness

For inherently risky situations with unforeseeable outcomes for gain or loss, such as investment decisions, trust is essential to use the advice of either human or technological advisors. Trust in automation is defined as the "attitude that an agent will help achieve an individual's goals in a situation characterized by uncertainty and vulnerability" (Lee & See, 2004, p. 51). Especially when a robo-advisor makes investment decisions or dynamically manages portfolios on its own, one is confronted with asymmetric dependencies and limited influence opportunities on the AI's decisions, which leads to a pronounced relevance of trust (Lewicki & Wiethoff, 2000). This attitude and related behaviors are vital to be studied to understand why, how and under which conditions AI advice is used.

Still, trust in AI is not automatically identical to trust in technology in general: While the latter is often judged on experience and functional characteristics, key factors such as ethics, transparency and explainability play a larger role when it comes to AI (Ashoori & Weisz, 2019; Hoff & Bashir, 2015). A helping theoretical guidance can be the Trustworthiness Assessement Model for AI (TrAM; Schlicker et al., 2025). This model describes the process of how the actual trustworthiness of an AI system is perceived, ultimately leading to trust and trusting behavior. Core of the model is the idea that there is a difference between the actual trustworthiness (AT) of an AI system, resulting from an alignment of the objectively given characteristics of the system and the individual standards for said system, and perceived trustworthiness (PT). Between AT and PT, there is a trustworthiness assessment process oriented on Funder's Realistic Accuracy Model (1995): The system provides certain cues for trustworthiness (e.g., accuracy information), that need to be detected and utilized by the user to build one's own PT. It can be different for each person, system and situation.

## 1.2 Trust signals as trust builders

A cue (also called *signal*), as understood by the theory, is an "information element that can be used to make a trust assessment about an agent" (Visser et al., 2014, p. 253). It can be actively used to achieve an appropriate level of trust. For this, different approaches have been taken (see Wischnewski et al., 2023 for an overview), such as increased transparency through, for example, explanations (XAI) (see Saeed & Omlin, 2023, for an overview) or literacy training (Colville & Ostern, 2024). While such interventions aimed to increase users' understanding or the systems' intelligibility have been shown to successfully assist in the trust calibration process, they inherently shift the burden of making the right judgments to the user.

## 1.3 Two alternative cues for trustworthiness: certificates and proofs through formal verification

In this study, we aim at investigating the effects of two very different potential cues for trustworthiness: certification and formal verification. In the following, we argue on the mechanisms of those signals that can be used to assess trustworthiness and in consequence lead to use of and compliance to AI advice.

### 1.3.1 Certificates

Originating from theoretical considerations such as trust and epistemic trust theory, signaling theory, and persuasion literature, an alternative approach to facilitate the trust calibration process is *certification* or labeling methods. Both approaches allow users to effortlessly gain insights into the trustworthiness of systems. *Signaling theory*, for example, focuses on the communication process of one party [here: developers or providers] to the other [here: end users]. Central in signaling theory is the assumption of an information asymmetry between both parties wherein one party is less informed (the end user) than the other (the developer or provider). Bridging this asymmetry, certifications or model labels can function as "signals which are actions that parties take to reveal their true type" (Kirmani & Rao, 2000, p. 66). In other words, certifications or model labels can not only convey information about the explicit properties of a model but also signal transparency intentions of the provider, possibly increasing end-users' trust, especially in AI (relating back to the specific importance on transparency and explainability, see chapter 1.1).

In the persuasion literature, in particular, the *elaboration likelihood model* (ELM) (Petty & Cacioppo, 1984) offers a more direct application of certifications and model labels as a means to communicate. At its core, the ELM describes how individuals process persuasive arguments by following either a central, effortful route of information processing, or a peripheral route of processing which requires less cognitive effort. Theoretically, certifications and model labels can function as cues that can effortlessly be processed via the peripheral route, informing users about important properties. However, processing via the central route is also possible if model labels induce deeper elaboration (Lowry et al., 2012), such as trade-offs between different properties. In sum, trust signals such as certifications and labels firmly ground the trust building process as a matter of communication (Liao & Sundar, 2022).

While such theoretical benefits exist, empirical work about their effectiveness is mixed. For example, Kim et al. (2008) found no effect of labels on trust, suggesting that participants in their study either lacked understanding or familiarity with the label's meaning. Likewise, in a more recent investigation, Wischnewski et al. (2024) found no effect of AI certification labels. The authors explain their results drawing on evidence from Adam et al. (2020) who introduce the trust tipping point: below a certain trustworthiness threshold, labels can effectively increase users' trust. However, with increased model trustworthiness, labels will not increase users' trust any further.

However, studies in the context of, for example, Web Assurance Seal Services (WASS) for e-commerce platforms could show that users' trust increased with the introduction of a label and users' concerns about the platforms were mitigated (Kim et al., 2016). A study on robo-advisors using two large survey data sets by Todd and Seay (2020) shows at least the subjective importance of designations/certification for investors, with around 71 % of investors with experience using robo-advisors indicating that they are very important when looking for a financial advisor.

### 1.3.2 Formal verification

The core of robo-advisors is that they can analyze investors' data that are collected by asking them questions as part of a risk profiling strategy (Bhatia et al., 2020). Still, as to the opaque nature of AI in general, laypeople and especially customers with limited information access cannot infer how and how well their provided data and investment preferences are considered by robo-advisors. This raises a broader question of reliability: how can we be confident that the AI system will behave as intended across the full range of possible user inputs? Two fundamentally different strategies address this challenge: *testing* and *formal verification*, offering very different levels of assurance.

Testing is the most common approach to quality assurance. Traditional testing checks whether a system behaves as intended for a number of specific test cases. While useful in practice, it has an inherent drawback: there is no certainty about how the AI system will behave during real-world operation. This is because testing evaluates the system on a finite test set, whereas in deployment, inputs arise from the entire large input space. As a result, it is almost impossible that a test scenario during development will exactly match the situations the system encounters during testing.

To tackle this shortcoming, *formal verification* reasons about the correctness of an AI system analytically on the entire space of potential inputs, independent of any (test) data set. This means, if an AI system is verified to be correct for its whole space of all potential inputs, then it is guaranteed to work correctly in all situations during its real-world operation. The result of this formal verification process is a *mathematical proof*, which guarantees the verified characteristics of an AI system.

In our case, verification could prove that for any user input, a robo-advisor recommends a low-risk portfolio when the user is risk-averse, and only recommends riskier portfolios to users who explicitly indicate they are comfortable with risk. If this property is verified, then it is guaranteed to hold in all real-world situations – regardless of which other (unusual) preferences the user might indicate. This is made possible by mathematical techniques such as constraint-based verification, where both the neural network and the desired property are encoded as a logical formula. For instance, the desired behavior could be expressed as:

$$\forall x: \text{RiskAverse}(x) \rightarrow \text{LowRisk}(f(x)) \land \text{RiskSeeking}(x) \rightarrow \text{PossiblyHighRisk}(f(x))$$

where x represents user input, the symbol $\forall$ indicates that x can be any point from our input space, the arrow $\rightarrow$ represents a logical implication, and $f(x)$ is a logical formula that models the input-output behavior of our AI system. A specialized solver then checks whether this formula is logically valid for all possible inputs. If it is, the system is proven correct across the entire input space and can be employed in real world safely; if not, the solver returns a concrete counterexample showing how the system could fail, requiring further development. Although verification is technically challenging and computationally intensive, it offers an absolute level of reliability—unlike testing.

People perceive formal verification as a more important seal for AI than those for transparency or trainings data and regard it as relevant for AI performance (Wischnewski et al., 2024). Hence, in our study, where a robo-advisor's performance is crucial for individual investment success, we expect positive effects of, especially, formal verification.

### 1.3.3 Hypotheses on the effects of certification and verification on trust

In this study, scenarios where no quality assurance measures are taken for a robo-advisor are compared to certified and verified advisors. The main difference between certification and verification is that certification claims certain characteristics or processes, while verification proofs them. Deduced from the previously presented research findings on certificates and mathematical verification as potential, yet differentially valuable cues for a robo-advisor's trustworthiness, we assume the following hypotheses:

*H1a:* The experienced trust in the artificially intelligent advisor increases with increasing levels of quality assurance of the advisor (none < certified < verified).

*H1b:* The experienced mistrust in the artificially intelligent advisor decreases with increasing levels of quality assurance of the advisor.

7Thus, we expect that an intransparent robo-advisor will lead to the lowest levels of trust and highest levels of mistrust, followed by a certified advisor and that a mathematically verified robo-advisor will elicit the highest level of trust and lowest level of mistrust. Exploratively, we want to additionally find out how the conditions compare to one where a human financial advisor is provided.

#### 1.3.4 Hypotheses on the effects of certification and verification on mental model precision and dependability

Additionally, the effects of different cues for trustworthiness on two more specific variables regarding the perception of the robo-advisor are regarded.

One is the *perceived precision of the mental model*. Mental models, also described as images (Johnson-Laird, 1980), are representations of the world or certain objects based on one's understanding of said object and one's knowledge (Johnson-Laird, 2010). They are essential for the general understanding of how a system, here a robo-advisor, operates (Johnson-Laird, 1980; Morris & Rouse, 1988). Endsley (2023) names the possession of adequate mental models of one another as a key requisite for teaming with intelligent systems. While objective model adequacy can be judged only after experiences with the robo-advisor, there is also a cognitive component to the individual judgement of adequacy. It can be framed as perceived precision of one's mental model and can be measured as one's conviction that one has knowledge on how a (technological) system works, what its use is and what its limitations are (Tausch, 2025). It is accordingly the subjective extent of knowledge on the typical mental model components (see, e.g., Carroll & Olson, 1988). This should be influenced by the amount and quality of cues offered for quality assurance, leading us to the following hypothesis:

*H1c:* The perceived precision of the mental model of the financial advisor increases with increasing level of quality assurance of the advisor.

The second variable is *dependability*, with its idea grounded in computing and later connected to the field of user experience research. Dependability can be defined as "the ability [of a computing system] to deliver service that can justifiably be trusted" (Avižienis et al., 2001, p. 1) or as the degree of a user feeling in control of the interaction with a system (Schrepp, 2023). It is included by Laugwitz et al. (2006) in their User Experience Questionnaire (UEQ) as a pragmatic product quality. Research has found that the usability of robo-advisors, of which dependability is a component, has a positive effect on the willingness to rely on it for one's wealth management (Yi et al., 2023). As such, we assume the following:

*H1d:* The perceived dependability of the financial advisor increases with increasing level of quality assurance of the advisor.

### 1.4 Trusting behavior

Trust in a certain technology has been identified as a central prerequisite to not only the general uptake of systems but also their appropriate use (Lee & See, 2004; Madhavan & Wiegmann, 2007; Parasuraman & Riley, 1997). To that end, users' trust should be appropriately calibrated to a system's actual trustworthiness (Wischnewski et al., 2023). A mismatch of the perceived and actual system trustworthiness can result in either undertrust (i.e., disuse - resistance to use the system) or overtrust (i.e., misuse - over-reliance on the system) (Parasuraman & Riley, 1997), both of which can result in serious consequences. What is actually the goal of actively incorporating signals for trustworthiness is to "optimize the accuracy with which people assess the trustworthiness of a system" to reach a state of "calibrated trusting behavior" (Schlicker et al., 2025, p. 14).

Generally, we know that trust in AI is central for acceptance and use of such technologies (Kelly et al., 2023): It influences, e.g., if users are willing to accept AI advice or decisions (Choudhury & Shamszare, 2023). A study on robo-advisors has shown that the willingness to use them depends, e.g., on financial experiences and expected emotions from their use (Hohenberger et al., 2019).

### 1.4.1 Hypotheses on investment as a form of trust behavior

Transferring knowledge on trusting behavior to the use of robo-advisors, we expect that trust behavior can be measured by observing how much money people are willing to provide a robo-advisor to invest for them. In the case of stable stakes (in this case by controlling them experimentally), investment of a lot of money would reflect high trusting behavior while hesitant or no investment would be a symptom of low behavioral trust. Again, it can be expected that the level of quality assurance of the robo-advisor will influence trust, also shown in concrete behavior.

*H2a:* The amount of money given to an artificially intelligent financial advisor increases with increasing level of quality assurance of the advisor.

Additionally, and according to the TrAM, we expect trust behavior to be a consequence of trust. Thus, we also assume an effect of trust, and a reverse effect of mistrust, on behavior, and that this effect mediates the effect of the level of quality assurance.

*H2b:* The amount of money given to a financial advisor increases with increasing trust (full mediation of level of quality assurance effect).

*H2c:* The amount of money given to a financial advisor decreases with increasing mistrust (full mediation of level of quality assurance effect).

### 1.5 Effects of calibration points on trust

An initial interaction with a robo-advisor will very much rely on attitudes and expectations, while continuous or repeated use will be dependent on the experiences that have been made with the system and the outcomes that are associated with its use. Within this process of interaction, trust is continuously calibrated – triggered by calibration points, such as trust violations, that affect the factors essential for trust (Hoff & Bashir, 2015). We know that trust in technologies is strongly coined by the experiences of the user and its perceived ability to perform (Lee & See, 2004). And while this is true for positive experiences such as successful investments of a robo-advisor, it seems to be even more pronounced with unfavorable events, as the impact of bad events and emotions overall is stronger than that of positive stimuli (Baumeister et al., 2001). Only one failure can lead to substantial losses in trust (Hoff & Bashir, 2015).

To investigate the effects of negative calibration points, we compare trust, mistrust and dependability rating following an either successful or failed investment:

*H3a:* When a financial advisor fails to invest participants' money, users trust the advisor less than when the advisor successfully invests participants' money.

*H3b:* When a financial advisor fails to invest participants' money, users mistrust the advisor more than when the advisor successfully invests participants' money.

*H3c:* When a financial advisor fails to invest participants' money, users evaluate the advisor's dependability worse than when the advisor successfully invests participants' money.

Additionally, we regard the within-person effects of especially successful investment, i.e., a positive calibration point, on (mis)trust and dependability changes:

*H4a:* Users' trust increases in all conditions after an investment success.

*H4b:* Users' mistrust decreases in all conditions after an investment success.

*H4c:* Users' perception of AI dependability increases in all conditions after an investment success.

## 1.6 Over-attribution and expectancy violation of trust signals

At least two limitations go in hand with the introduction of trust signals. On the one hand, these signals might be misused by malicious agents who misrepresent the true nature of a system with the aim of deception. To overcome this possibility, Liao and Sundar (2022) suggest solutions from evolutionary psychology, arguing that humans have developed to favor signals or cues that are reliable and costly. Thus, mathematical verification, being computationally expensive and time-consuming, should therefore be the least prone to misuse.

On the other hand, users might misunderstand signals and cues. For example, early investigations of the effectiveness of online web assurance seals found that users commonly over-relied on and misinterpreted assurances (Kirlappos et al., 2012; Turow, 2023). Moreover, trust signals might also evoke wrong expectations. Again, relying on research that examined online purchasing in the early days of the internet, Houston and Taylor (1999) found what they called the *expectation gap*. In their empirical work, the authors noticed that a WebTrust$^{SM}$ assurance seal did not communicate the intended purpose (assurance of business and security practices) but instead raised consumers' product quality expectations.

In turn, not meeting expectations due to misattributions of a trust signal's functionality might lead to skepticism and mistrust, possibly undermining the original intended purpose of a trust signal (Lins et al., 2024). This is decisive for two reasons: (1) For any trust signal, but especially for more complex ones open to misinterpretation such as mathematical verification, it is essential to raise correct expectations to make them valuable trustworthiness cues; (2) when a calibration point for trust is reached that can question a system's integrity, wrong expectations due to misattributions can lead to substantially lower trust and trusting behavior.

### 1.6.1 Backlash effects of level of quality assurance after failed investments

Based on those insights, a moderating effect of the level of quality assurance on the effects of an unsuccessful investment of a robo-advisor can be expected. If an agent is certified or even mathematically proven, the negative calibration point should influence (mis)trust and dependability more than if it is intransparent, due to over-attribution and distorted feelings of safety due to promises of certificates or verification.

*H5a:* Users' trust reduces more after a failed investment success, the higher the level of quality assurance of the financial advisor.

*H5b:* Users' mistrust increases more after a failed investment success, the higher the level of quality assurance of the financial advisor.

*H5c:* Users' perception of dependability reduces more after a failed investment success, the higher the level of quality assurance of the financial advisor.

## 2   Methods: Online Experiment

Using an online-experimental setting, participants were confronted with a scenario in which they were able to invest money via an online financial advisor. Presenting textual and visual vignettes, the financial advisor and the success of the investment were manipulated to examine their effect on trust, trusting behavior and investments. This was done in a 4x2 factorial mixed design, with the between-subjects factors *level of quality assurance of the advisor* (none, certified or verified robo-advisor + human advisor [control condition]), and *success of the advisor* (successful, unsuccessful). Participants were partially randomized, maintaining equal distribution over the conditions based on fully finished questionnaires and an equal gender balance.

The study was pre-registered at OSF https://osf.io/2gudn/ (including all presented hypotheses and additional analyses that are not considered in this paper) and received an ethics approval with reference number 919 from the local ethics committee.

### 2.1   Study design and vignettes

The study was conducted as an online study via the SoSci Survey platform with a German-based Prolific sample from December 17, 2024 to January 6, 2025. Participants were randomly assigned to one of the four experimental conditions concerning the advisor before starting to follow the experimental procedure depicted in Figure 1. This procedure contained text and image vignettes describing the different financial advisors, while the overall scenario, the simulated counselling interview and the portfolio description were held stable across all conditions.

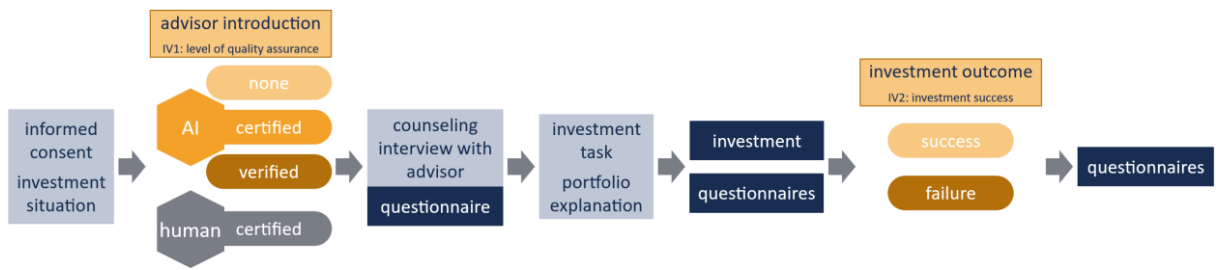

*Figure 1. Experimental procedure and manipulations.*

Participants were asked to imagine a scenario where they inherited 100,000 € and decided to use an advisor of the online bank FinAid to help invest parts of the money. Then, the respective advisor was introduced with an image, a text describing its general characteristics (see

Table 1 for an English translation of the originally German vignette texts and images) and a simulated screenshot of the websites giving details on the respective level of quality assurance of the advisor.

*Table 1. Presentation of the different financial advisors and their respective introductory texts.*

| Advisor | FinBot intrasparent | FinBot certified | FinBot proven | Holger Müller Certified |
|---|---|---|---|---|
| Short summary* | No information on quality assurance | Certification label from TÜV AI.Lab** for safe and ethical AI solutions | Formal verification of reliability and correct results considering client's risk preferences and investment goals | Structurally trained and certified by IHK to fulfill standards in his work processes |
| Picture of the advisor shown to participants | 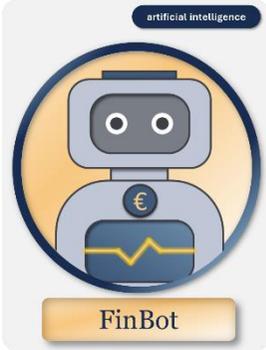 | 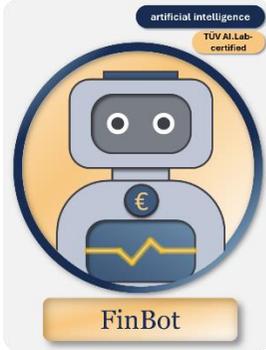 | 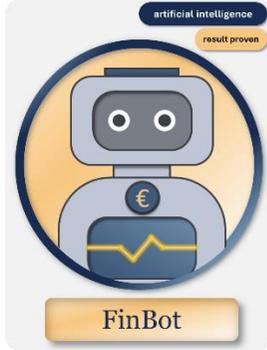 | 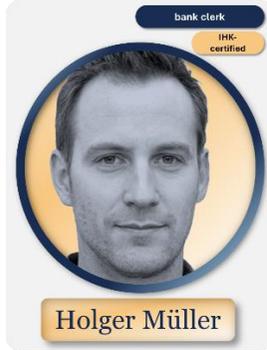 |
| Introduction of advisor | / | The AI investment advisor assigned to you is called FinBot. FinBot is trained as an investment algorithm and has been used by FinAid to provide advice since 2022. | | The investment advisor assigned to you is Holger Müller. He trained as a banker and has been working as an advisor at FinAid since 2022. |
| Further information on advisor | / | FinBot is an AI developed within the company that has been certified by TÜV, i.e. tested by an external body. This ensures that the AI meets the necessary standards for its work and follows appropriate processes in its work. However, this does not ensure that FinBot functions correctly. You can view the criteria for the certification of FinBot below. | FinBot is an AI developed in the company that has been mathematically verified. Its correct functioning has been formally proven, so its results have been computer-checked. This ensures that the AI reliably delivers the results it is supposed to deliver. There are thus formal guarantees for its correct functioning. However, this does not ensure standards and appropriate processes in the work of FinBot. You can view the relevant information on the evidence below. | The bank advisors have been trained in the company in accordance with the standards of the Chamber of Industry and Commerce and have been externally audited. This ensures that they meet the necessary standards for their work and follow appropriate processes in their work. However, this does not ensure that the financial advisors work correctly in general. You can view the criteria for the certification of our bank advisors below. |

*Note. The short summary was not included in the study, it is only used here for illustrative purposes.

** Note. The TÜV is a well-known German certification agency (renowned for checking cars for safety). Its AI.Lab is an actually existing department within the TÜV, but is not yet certifying AI systems. Still, it was used for authenticity and familiarity in this study as the certifying agency.

13Afterwards, a counselling interview with the advisor asked for the financial experiences and investment preferences of the participants. The data is mostly irrelevant for the study (except for sociodemographics), but important for immersion to underline the importance of mathematically proving that this information is actually considered during the investment.

The investment situation is then introduced informing participants that they can invest any amount between 0 and 100,000 € with the advisor, with the alternative of the non-invested money being put into a call money account associated with inflation losses and costs. The advisor presents the portfolio that is individually and dynamically complied from investments in government bonds, real estate funds, shares and cryptocurrencies, representing a balance of high- and low-risk investment opportunities. After being asked how much money participants give to the advisor to be invested, a questionnaire battery is answered.

Another vignette element in the end either tells the participants that, after five years of investing, the advisor could realize 1) a gain of 10% per year or 2) a loss of 3% per year. Again, participants were randomly assigned to one condition each. A (un)favorable market development and an (un)successful investment strategy were named as reasons for success or failure. Afterwards, willingness to invest further as well as a repetition of some outcomes were measured to regard dynamics on the within-person level.

## 2.2 Measurement Instruments

We used a number of questionnaire instruments and single item measures, which are summarized in Table 2. Additionally, we assessed investment experience and preferences, sociodemographics and readiness for digitalization, which are not essential for the following analyses. A manipulation check asked participants who combined and managed their portfolio.

Global trust and mistrust were measured both before the investment and after information on the success of the investment after an imaginary time period of five years. This post measurement also included advisor dependability.

*Table 2. Constructs and measurements used in the study.*

| Construct | Source | Items | Scale | Example item |
|---|---|---|---|---|
| Mental Model of Collaborative Technologies Scale | Tausch (2025) | 7 | 1 to 5 | I have an idea of how [the system] works. |
| Investment | Newly developed | 1 | 0 to 100,000 € | How much of your in total 1000.000€ will you give to [your advisor] to invest it for you? |
| Investment after success information | Newly developed | 1 | 1 to 5 | I give this much money to the [advisor] for further investments: [no money to all money] |
| Portfolio risk assessment | Newly developed | 1 | 1 to 11 | How risky do you perceive the presented portfolio, independent of the performance of the advisor? |
| Advisor dependability | UEQ by Laugwitz et al. (2006) | 4 | 1 to 7 | Please indicate your assessment of [your advisor]: Unpredictable … predictable |
| Trust - global | Trust in AI Scale by Wischnewski et al. (2025) | 5 | 1 to 5 | I can depend on [the advisor]. |
| Trust – integrity | | 5 | | |
| Trust – unbiasedness | | 5 | | |
| Trust – transparency | | 5 | | |
| Trust - ability | | 5 | | |
| Trust - vigilance | | 5 | | |
| Trust in different instances | Newly developed | 11 | 1 to 8 | How much do you trust the following instances and processes? [certification-instances such as the TÜV] |
| Perceived control | Menon Empowerment Scale | 3 | 1 to 6 | I can influence the way investments are done in the financial advisory process. |

Note: The questionnaires were used in German. The items have been translated for this table.

## 2.3 Sample

533 people took part in the study running from December 17, 2024, to January 6, 2025 with recruitment via the Platform Prolific. Participants received a compensation of 3,07 €. After excluding participants who, according to the preregistration, did not fully finish the questionnaire (6) or failed in the control item (modified to a more subtle solution after preregistration, 7), a sample size of 520 was achieved. This is sufficient based on the necessary sample size of 489 participants for an effect of $f = 0.15$ and a power of .80 on the standard α-error level of .05, calculated using G*Power.

The sample contained 256 male and 255 female people, 9 chose not to assign. They were 35.32 (11.59) years old at average. A share of 53.3 % were working or employed on a full-time base, 18.7 % in part-time, 16.9 % were studying, going to school or learning a job and 11.2 % were unemployed. Participants were distributed to the conditions as can be seen in Table 3.

*Table 3. Distribution of participants to the experimental conditions.*

| | No quality assurance | certified | verified | human | all |
|---|---|---|---|---|---|
| Successful investment | 65 | 70 | 60 | 65 | 260 |
| Failed investment | 64 | 60 | 73 | 63 | 260 |
| all | 129 | 130 | 133 | 128 | 520 |

## 3 Results

### 3.1 Descriptives

**Error! Reference source not found.** contains all major descriptives for the measured variables, including means with standard deviations, scale reliabilities and correlations. For the investment, participants gave on average 55,454 € to the advisors, with a median of 60,000 € and the mode, i.e., the most common value, being 50,000 €.

*Table 4. Scale measures, means, standard deviations and correlations of the central constructs.*

| | scale | α | M | SD | r | | | | | | | | | |
|---|---|---|---|---|---|---|---|---|---|---|---|---|---|---|
| | | | | | 1 | 2 | 3 | 4 | 5 | 6 | 7 | 8 | 9 | 10 |
| 1. investment (€) | 0 – 100,000 | / | 55,455.00 | 31,359.75 | 1 | | | | | | | | | |
| 2. global trust | 1-5 | 0.92 | 3.13 | 0.87 | .46** | 1 | | | | | | | | |
| 3. global trust (post measure) | 1-5 | 0.90 | 2.99 | 0.94 | .30** | .63** | 1 | | | | | | | |
| 4. mistrust | 1-5 | 0.89 | 3.95 | 0.73 | -.14** | -.27** | -.23** | 1 | | | | | | |
| 5. mistrust (post measure) | 1-5 | 0.92 | 3.96 | 0.82 | -.12** | -.24** | -.47** | .75** | 1 | | | | | |
| 6. precision of mental model | 1-5 | 0.84 | 3.17 | 0.73 | .17** | .35** | .29** | -.11** | -.11** | 1 | | | | |
| 7. dependability | 1-7 | 0.83 | 4.50 | 1.05 | .45** | .75** | .48** | -.23** | -.19** | .30** | 1 | | | |
| 8. dependability (post measure) | 1-7 | 0.89 | 4.16 | 1.33 | .29** | .51** | .85** | -.23** | -.48** | .26** | .54** | 1 | | |
| 9. investment risk | 1-11 | / | 6.35 | 2.07 | -.30** | -.48** | -.33** | .10* | .09* | -.13** | -.53** | -.33** | 1 | |
| 10. control experience | 1-6 | 0.94 | 3.73 | 1.23 | .12** | .40** | .24** | -.10** | -.07 | .21** | .30** | .18** | -.20** | 1 |

Note. The correlations are Pearson correlations tested in a one-sided manner.

\* p ≤ .05; ** p ≤ .01

## 3.2 Hypotheses testing

Before analyzing the hypotheses, assumptions were tested using Levene's tests indicating significant deviations from homogeneity of variances for H1c, H4a, H3c and H4c. However, given the robustness of the applied methods to such violations, we proceeded with the analyses as planned.

### 3.2.1 Main effects of level of quality assurance

The first group of hypotheses tests was considered with the differences between the three experimental conditions and the central outcome variables (mis)trust, perceived precision of the mental model, and advisor dependability rating. An overview comparing all conditions can be found in Table 5.

*Table 5. Means, standard deviations and ANOVA results comparing the three experimental conditions.*

|  | AI with no quality assurance | | AI certified | | AI verified | | human advisor | | comparison between AI conditions (without human) | | | |
|---|---|---|---|---|---|---|---|---|---|---|---|---|
|  | M | SD | M | SD | M | SD | M | SD | F | df | p | $\eta^2$ |
| Experienced trust | 3.16 | 0.84 | 3.06 | 0.85 | 3.04 | 0.96 | 3.27 | 0.82 | 0.66 | 2, 389 | .580 | .003 |
| Experienced mistrust | 3.97 | 0.72 | 3.93 | 0.74 | 3.92 | 0.78 | 4.00 | 0.66 | 0.12 | 2, 389 | .886 | .001 |
| Precision of the mental model | 2.95 | 0.77 | 3.16 | 0.75 | 3.3 | 0.64 | 3.25 | 0.71 | 7.78 | 2, 389 | <.001 | .038 |
| Perceived dependability | 4.43 | 1 | 4.45 | 1.05 | 4.41 | 1.06 | 4.72 | 0.97 | 0.06 | 2, 389 | .944 | .000 |

The analysis was not significant for *H1a* investigating trust, $F(2, 389) = 0.66$, $p = .520$, and for *H1b* investigating mistrust, $F(2, 398) = 0.12$, $p = .886$, $\eta^2 = 0.001$, which demonstrates that the perceived trust and mistrust do not significantly change with increasing level of quality assurance of the advisor and that *H1a* and *b* cannot be accepted.

The ANOVA using the mental model to test *H1c* showed significant results, $F(2, 391) = 7.78$, $p < .001$, $\eta^2 = .038$, meaning the participants' perceived precision of the mental model of the financial advisor increases with increasing level of quality assurance. Post-hoc comparisons using Tukey's test revealed a significant difference for the mean between the non-quality-assured and the verified advisor (*see* Table 5*), p < .001. H1c* can thus be accepted.

The ANOVA on dependability was not significant, $F(2, 391) = 0.06$, $p = .944$, $\eta^2 < .001$, suggesting that the perceived dependability of the artificially intelligent advisor does not increase with increasing level of quality assurance. Thus, *H1d* has to be rejected.

### 3.2.2 Investment Height

The central hypothesis *H2a* regards the difference of money given to the financial advisor for investment between the experimental conditions. The one-way ANOVA comparing the conditions' investments was not significant, $F(2, 391) = 0.62$, $p = .538$, $\eta^2 = 0.003$, indicating that the money given to an artificial advisor does not increase with increasing level of quality assurance of the AI advisor.

Descriptively though, the investment varied substantially between the conditions, especially when regarding mode and median, as can be seen in Figure 2 and Table 6.

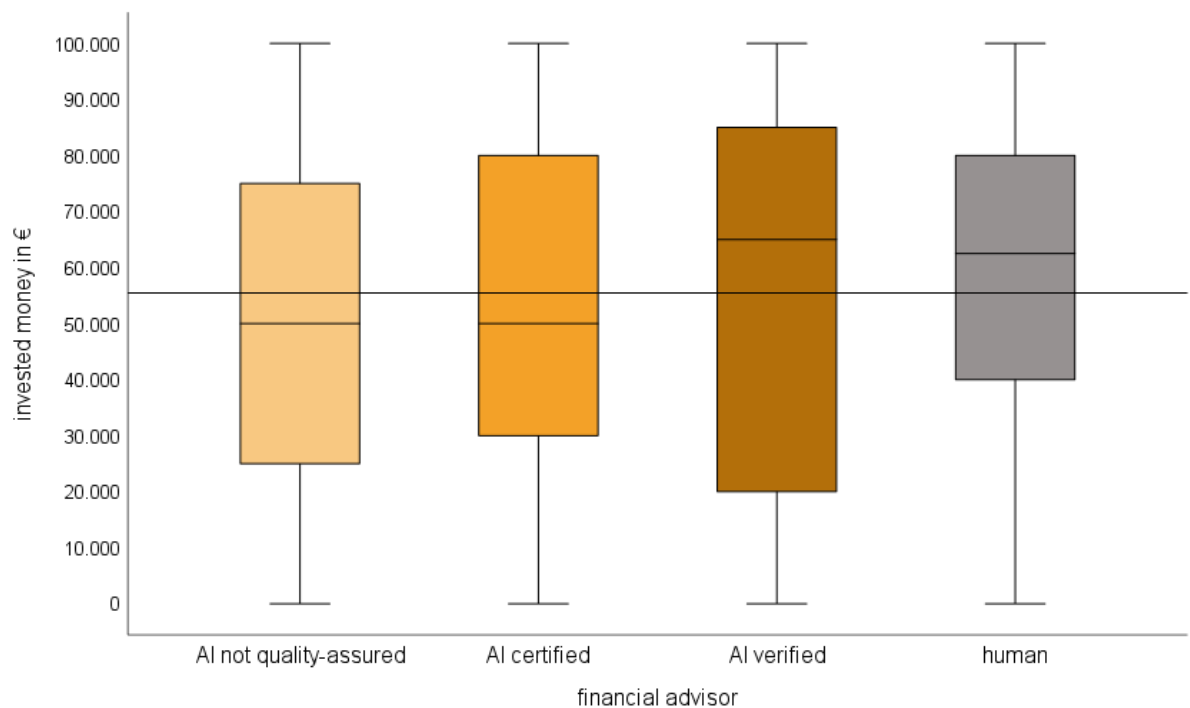

Figure 2. Boxplot showing the median investment in € in the four conditions as well as the overall mean investment.

Table 6. Descriptives for the investment in the four experimental conditions.

|  | advisor | | | | |
| --- | --- | --- | --- | --- | --- |
|  | overall | not quality-assured | certified | verified | human |
| M (SD) | 55,454 | 51,673.13 | 54,454.23 | 56,019.55 | 59,696.21 |
| median | 60,000 | 50,000 | 50,000 | 65,000 | 62,500 |
| mode | 50,000 | 50,000 | 50,000 | 80,000 | 50,000 |
| % of 0 € investment (%) | 5.6% | 5.4% | 6.2% | 7.5% | 3.1% |
| % of 100.000 € investment | 11% | 9.3% | 1.5% | 12.8% | 12.5% |

Testing the investment for normality, it can be found that in none of the conditions, normality is given (KS-test's $p < .001$) and that the frequencies of different investment values are also visually overall not normally distributed (Figure 3).

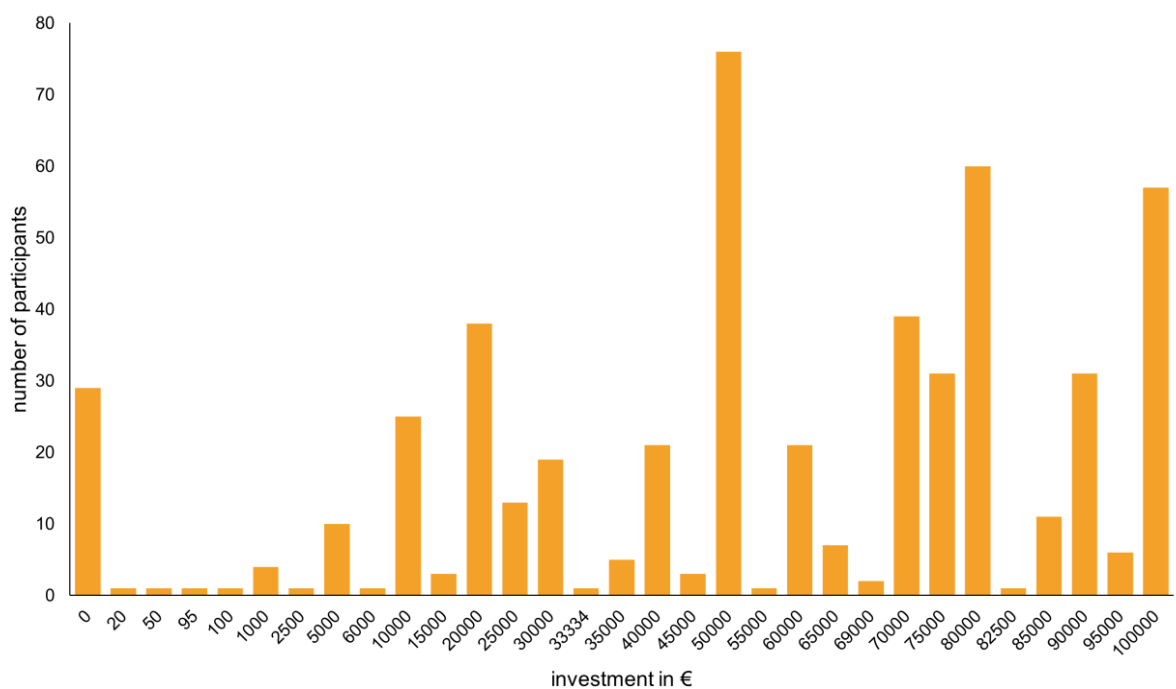

*Figure 3. Frequency plot of investment sums.*

For this reason, an additional Kruskal-Wallis Test was chosen as a non-parametric group comparison alternative based on the median of the subsamples. The mean ranks were 186.99 for the intransparent AI, 196.80 for the certified and 205.43 for the proven AI, $\chi^2(2) = 1.75$, $p = .417$. Thus, H2a cannot be accepted based on group comparisons, despite descriptive evidence supporting the hypothesis.

Exploratively, also the investment in the AI vs. the human advisors were compared. A t-test comparing the money invested for all AI conditions taken together with the human advisor condition yielded significant results, $t(255) = -2.14$, $p = .017$, $d = -0.27$.

*3.2.2.1 Mediation effects of trust and mistrust on investment*

A regression analysis of trust on investment was conducted, $F(1, 389) = 123.25$, $p < .001$, explaining 24 % of the variance. This shows that the amount of money given to a financial advisor increases with increasing trust, supporting *H2b*. However, the mediation model, $R^2 = .25$, $F(2, 389) = 63.75$, $p = .000$, does not profit from adding the level of quality assurance as an variable, $b = 3201.05$, $p = .271$. This stays in line with the ANOVA analyses.

At the same time, the amount of money given decreases significantly with increasing mistrust, $F(1, 389) = 9.81$, $p = .002$. The model explains 3 % of the variance. Again, the mediation effect of level of quality assurance was not given with it not predicting mistrust, $R^2 = .001$, $F(1,390) = 0.23$, $p = .633$ and thus not contributing to the overall model to explain investment height, $b = 2023.79$, $p = .300$. *H2b* and *c* can therefore partially be accepted for the link between trust and investment behavior, but with a larger effect for trust than for mistrust. No mediation effect is observed.

### 3.2.3 Main effects of failure of advisor

An independent samples t-test compared the reported trust after a failed or successful investment: It showed significant difference for both trust, $t(518) = 14.21$, $p < .001$, $d = 1.25$ and mistrust, $t(518) = -7.5$, $p < .001$, $d = -.66$. This suggests that the participants had significantly more trust as well as less mistrust for the advisor after a successful investment rather than a failed investment over all conditions, thus supporting hypotheses *H3a* and *H3b*. A t-test (without given homogeneity of variances) comparing also the users' evaluation of dependability after the investments were either successful or not, was significant, $t(502.86) = 15.46$, $p < .001$, $d = 1.36$. The mean dependability rating after a successful investment was 4.91 (1.00), while it was 3.42 (1.19) after a failed investment. This supports *H3c*.

### 3.2.4 Change in (mis)trust and dependability after investment success

In addition to group comparisons, measures before and after success information were compared. An ANOVA with repeated measures confirms that trust increases, $F(1, 256) = 104.55$, $p < .001$, $\eta^2 = .29$, and mistrust decreases, $F(1, 256) = 46.6$, $p < .001$, $\eta^2 = .15$, after an investment success in all conditions, leading to the acceptance of *H4a* and *H4b*. Also the repeated measures ANOVA on dependability brought up significant differences between pre and post measurement, $F(1, 256) = 70.3$, $p < .001$, $\eta^2 = .22$. Users' perception of dependability increased after the investment was successful, supporting hypothesis *H4c*.

### 3.2.5 Interaction effect of level of quality assurance and investment success

A general linear model was performed to investigate the interaction between level of quality assurance and trust (*H5a*), mistrust (*H5b*) and advisor dependability (*H5c*) changes after failed investments. The effect of time on trust, $\lambda = 0.63$, $F(1, 256) = 149.77$, $p < .001$, $\eta^2 = .37$, and the interaction effect of time and level of quality assurance / advisor, $\lambda = 0.03$, $F(3, 256) = 2.85$, $p = .038$, $\eta^2 = .03$, are both significant. A post-hoc Tukey test reveals no differences between the different advisors, p > .05. A graphical representation of the interaction can be found in Figure 4

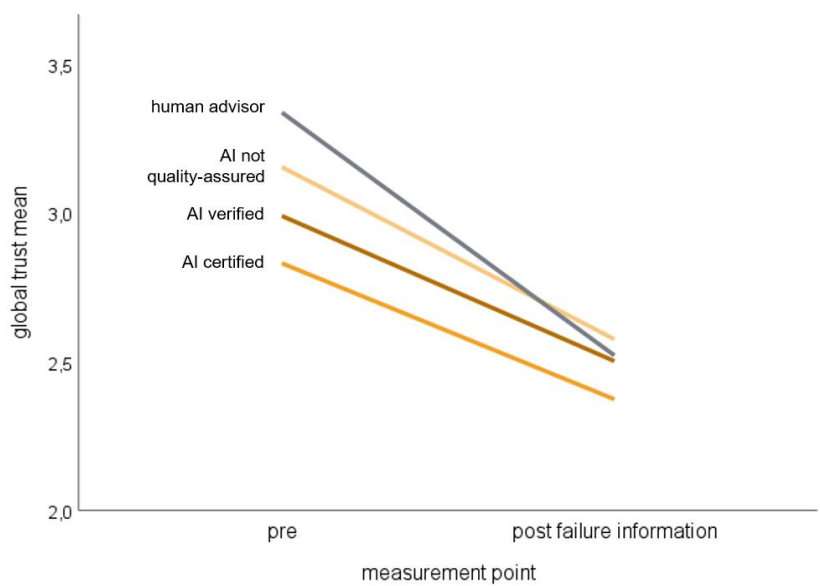

*Figure 4. Change in global trust after investment failure information in the four conditions.*

For mistrust, only the effect of time is significant, $p < .001$, but not the interaction effect of time and level of quality assurance / advisor, $p > .05$. The same holds for dependability, so that neither hypothesis *H4a*, nor *H4b* or *H4c* are clearly supported by the data.

## 4 Discussion

Summarizing the results, the level of quality assurance of the advisor had surprisingly near to no effect of any of our outcome variables, except for people's perception of their own mental model of the advisor. Though interesting correlations between constructs relevant for technology use and acceptance were found, the main finding is that overall, the information that were presented about the robo-advisor seem to only play a minor role beyond the mental model, although they were of substantially different value for assessing the advisor's quality. Still, when measuring behavior in the form of a sum given to the advisor for investments, we find that it a) makes a difference if the advisor is a human and b) it seems to also matter if the AI is verified to yield the highest behavioral trust. With the first result being statistically significant, the different investment behavior can only be seen in the descriptive data – still there, it is quite obvious that people are investing on average about 2,000 € and in median 15,000 € more when confronted with a mathematically verified robo-advisor.

Regarding the subjective data, there are no differences in trust or mistrust in the three robo-advisors as well as no differences in their rated dependability. The only difference (with a small effect) was in the perceived clarity of people's mental model of the AI, which was higher when the AI was verified than when it was not quality-assured by any means.

We find that while trust explains about a quarter of the variance of trusting behavior, i.e., the amount of money invested, it takes no mediating role between different levels of quality assurance of a robo-advisor and the money given to it for investment. Mistrust seems to not produce reverse effects, but has only small, yet negative, effects on investment height.

Additionally, the effects of a potentially trust-calibrating event were examined by giving participants feedback on either the advisor's successful or failed investments after a certain time period. As was expected, trust and advisor's dependability differ largely between those with a successful and with a failed investment, and mistrust at a medium effect size. Comparing before and after-measurements, we find that trust and dependability increase with a large effect after advisor success information (although it was explicitly stated that this was also due to market developments), while mistrust decreases, not to the same extent, yet also with a large effect.

Results regarding the interaction of the advisor's testing and success or failure information remain unclear. While there is clearly no interaction regarding mistrust and dependability rating, there is an unclear, yet small, effect regarding trust, suggesting that maybe trust towards the human, while starting the highest, suffers the most from a failed investment. An overview over the hypotheses and their confirmation or rejection can be found in Figure 5.

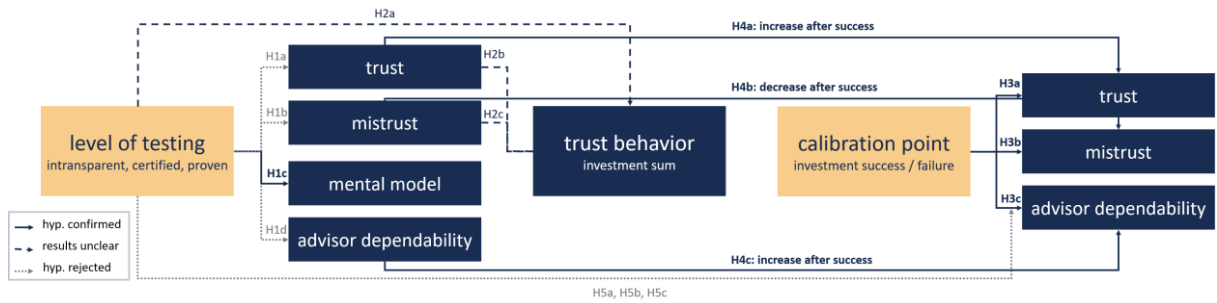

*Figure 5.*

## 4.1 A clearer mental model via quality assurance measures

First of all, the study shows that the more reliably quality assured an AI system is, the clearer is the mental model of abilities and cooperation opportunities of the robo-advisor. Descriptively, the mental model of the verified AI is perceived similarly precise as that of the human financial advisor – which is a clear sign that people seem to have a good image in mind of how to use the robo-advisor. The importance of mental models shows in them being an own reality: People stick to them, sometimes without reflection of accuracy or completeness, and orient their expectations or reasoning on them (Westbrook, 2006). If no additional information or quality assurance is given, people are left with an unclear mental model, allowing for misinterpretations of AI behavior or expectancy violations. The goal should be to help people achieve a mental model that is optimally synchronized with objective system characteristics, so that opportunities and limitations can be assessed appropriately. As Westbrook (2006) argues, system designers have to get an understanding of the key characteristics for end-users to build their mental models: We see that formal verification might be such a key feature. It can help speed up a process of mental model formation that would otherwise take longer interactions and multiple feedback loops, adjusting the model with own experience (Westbrook, 2006).

## 4.2 Explaining the missing effect of level of quality assurance and addressing limitations

As stated above, the expected effects of the cues for quality assurance that were provided in this study, a process certificate and a formal verification, did not lead to enhanced trust or perceived or decreased mistrust. There can be multiple explanations for that. The most probable one lies in the process of cue use. Funder's realistic Accuracy Model and TrAM already suggest that the pure availability of cues for trustworthiness does not automatically lead to their (appropriate) use. Describing and visualizing the certificate / the formal verification, we made sure that those cues were both relevant (with a higher relevance of formal verification) and easily available for incorporating them into robo-advisor judgement. What we were not able to control is if and how those cues were detected and utilized (see Funder, 1995). It could be that people rated the information on certification and verification as unimportant or that they based their trust and decision on something else such as common tendencies towards certain numbers / percentages (we see more "favored" investment options like 0, 20.000, 50.000 and 80.000 €) or their trust in the online bank or financial system. As such, our results are not standalone: From a recent study, we can find that giving people a seal guaranteeing mathematical proof of an AI does not lead to higher trust (Wischnewski et al., 2024). This counter-hypothesized result can be explained, e.g., by limited understandability and salience of seals, but also the limited potentials of seals to only leverage trust to a certain threshold (see Wischnewski et al., 2024). Especially for formal verification, limited understandability can be a problem, as it is not a commonly known concept, is mathematically complex and has not been extensively explained in the study material.

Another reason for quality assurance not playing a role for trust and dependability rating can be the study design using an imaginative scenario and pretended behavior. Although we tried to come close to actual behavior by asking for a sum to invest with the advisor and by presenting an immersive scenario using website visualizations and an imaginary chat with the advisor, the situation still was very different from being real. People inherited the imaginary money and thus it was not their own, the scenario and investment did not impact their actual financial status in the moment or in the future and there was nothing to really win or lose. The missing stakes of the decision are generally a problem of the study: As trust definitions and theories suggest, stakes are essential for trust to play a role. The study by Ashoori and Weisz (2019) shows that trust in AI depends heavily of the stakes of a decision: While AI was mostly accepted for low-stake situations, skepticism towards it grew with rising stakes. All those discrepancies to a field experiment might have led to trust and investment behavior being perceived as less decisive and potentially the quality of the advisor not playing an essential role for behavior within the experiment. Still, finding effects of trust comparing failed and successful investments suggests that also our imaginary manipulations worked and that finding effects was generally possible within our vignette design.

The unclear results for investment behavior are especially interesting: Even if no significant effect can be found, there are strong descriptive hints towards effects of quality assurance levels. Our free-text answer format, enabling participants to type in any number from 0 to 100.000 €, resulted in a high variability of answers and an uneven distribution of investments sums. This could have covered descriptively visible differences. In the non-quality assured condition, the lowest amount invested was 5,000 €, whilst for the other conditions, it was 50 or 100 €, potentially distorting the results. While this argues in favor of existing investment differences, it also opens up a new question: Why are subjectively reported trust and dependability not affected by quality assurance measures, while behavior is?

One assumption is that there is not a direct, linear transfer of trust to behavior. Studies show that factors such as behavioral intention or self-efficacy also influence compliance behavior beyond trust (Lowry et al., 2014), and the Knowledge Behavior Gap Model (Stibe et al., 2022) suggests that there are vital steps between knowledge (e.g., about the quality of an AI solution) and behavior, subsumed as acceptance and intention. The other assumption is that seals potentially do not affect cognition on a conscious level, but rather affect behavior unconsciously. This would mean that when being asked to give information about one's trust, seals do not show to be effective, but they are reflected indirectly in trusting behavior, without people even realizing. As to the ambiguity of our results, this aspect needs further exploration (see 4.5).

### 4.3 Comparison between human and robo-advisors

Consistent with the findings of the study by Hodge et al. (2021), we can find no difference in preference for a human vs. a robot financial advisor. Whilst Hodge's study investigated the likelihood of relying on the advisor, we looked at trust and investment behavior. Still Hodge et al. (2021) find that there is an interaction effect of advisor type and its humanization via a name: Under complex conditions, as in our experiments, people prefer a named human advisor, but an unnamed robo-advisor. In our experiment, both entities were given a name, yet the robo-advisor was called FinBot, which is not implying high levels of humanization. Thus, our results are in line with Hodge's findings and suggest great potential for robo-advisors beyond their current use in the financial market. A little caution when interpreting the results has to be used as our investment behavior was not real, i.e., not using people's own money and not having real-world consequences for their financial status. It could be that the willingness to invest one's own money, especially when it is not inherited and hardly owned or urgently needed, would still be higher with a human financial advisor as to its greater familiarity and thus basic level of trust. For example Ashoori and Weisz (2019) showed that decisions were perceived more trustworthy when finally made by the human, not AI, despite AI's perception as being more objective. Still, under our conditions, we could not find this effect of preference of human advisors, except for a slightly higher dependability perception.

### 4.4 The effect of success and failure information

While the effects of our quality assurance measures were insignificant, the effects of investment success or failure were medium to large. Even though all information was equal except for the gain of 10 % p.a. or loss of 3% p.a. of the investment, and despite the explicit reminder that those outcomes are due to a periodically unsuccessful investment strategy and unfavorable market conditions, trust and dependability were substantially increasing and mistrust decreasing with successful investment. Failure lead to reverse effects. As such, the investment outcomes seem to be rated as a much more important cue for the AI's trustworthiness (at least on a conscious level) than information on its certification or verification. This is problematic because concerning informational value, certification and especially formal verification are more important and more directly associated with a robo-advisor's overall quality, while investment success is influenced by many things beyond quality. Thus, it might lead to over- or undertrust in relation to the actual performance potential of a robo-advisor.

### 4.5 Research & theory implications

The interpretation of unconscious influences of seals on behavior, yet not on subjective trust, would mean that many of the studies we have on trust, attitudes and behavioral intentions might leave a knowledge gap. If there is a subconscious processing of cues for trustworthiness, it needs to be examined using different methods than self-report questionnaires and using more behavioral measures combined, e.g., with physiological data and methods such as eye-tracking. As it is only one potential interpretation of our results, which are also quite ambiguous, future studies should thoroughly investigate both subjective and objective, behavioral variables and compare the results. Models like the TrAM should then elaborate the interface between trust and trusting behavior, as we need more concrete process configurations to understand communalities and gaps between the two.

We ourselves plan on refining the paradigm of this study to clarify our results: The study we are currently developing includes an incentive system for simulating stakes of the investment decision, a different measurement of investment behavior that leaves less room for uncontrollable variation of input values, and a prolonged, more transparent explanation of formal verification to ensure an understanding of the meaning and its implications. We hope that by this, especially the ambiguous results concerning investment behavior and formal verification are showing a clearer direction.

In addition to this, it would be of great interest to not only target laypeople or end-users of AI systems and their reaction to quality assurance measures, but also professionals, e.g., purchasing robo-advisors for banking houses. Potentially their greater expertise and more in-depth understanding of the importance of quality assurance will lead to different effects and might prove that formal verification is not only a valuable, but also an effectively used cue for trust and AI use behavior.

### 4.6 Practical Implications

The central practical implication is that, while formal verification is essential in providing high-quality assurance of algorithmic results, it might not be a valuable clue for non-expert end-users. Finding no subjective benefits of it and only tendencies towards higher trusting behavior, we can state that the expenses for conducting formal verification should 1) be undertaken only when the benefit is very clear and when 2) its processes and results are edited carefully and user-friendly to allow for its understanding and appropriate detection and use. This could especially be the case when providers of AI systems are communicating with customers such as Chief AI officers, purchasing agents specialized in software who are willing and able to invest in understanding the true value of formal verification for buying and using AI that reliably ensures its results.

Another practical learning from our results concerns the role of robo-advisor investment success. We can clearly see that the success is (partially) mistaken as a cue for trustworthiness, leading to a peak in trust and dependability rating. While this can be a positive result from the viewpoint of companies offering robo-advisor services, society and consumer protection need to be aware of those mechanisms and develop counteracting measures preventing overtrust and increasingly risky investment decisions, especially when the robo-advisor quality has not been ensured by reliable measures such as thorough certification and formal verification. On the other hand, measures need to be taken to avoid undertrust following a negative calibration point in the form of loss or unsuccessful investments. If caused not by inappropriate AI investment behavior, it has to be made clear to consumers that market conditions are volatile and can, despite sensible investment strategies, vastly influence financial outcomes of investment activities. A potential measure to counteract those undertrust effects could be to present information side-by-side showing how other forms of investments (e.g., non-AI based investments) are performing in the same period and giving contextual information on economic development at the calibration point.

# 5 References

Adam, M., Niehage, L., Lins, S, Benlian, A, & Sunyaev, A (2020). Stumbling over the trust tipping point–The effectiveness of web seals at different levels of website trustworthiness. In Association for Information Systems (Chair), *European Conference on Information Systems,* Marrakech, Morocco.

Albarghouthi, A. (2021). *Introduction to neural network verification.* https://verifieddeeplearning.com/

Ashoori, M., & Weisz, J. D. (2019). *In AI We Trust? Factors That Influence Trustworthiness of AI-infused Decision-Making Processes.* https://doi.org/10.48550/arXiv.1912.02675

Avižienis, A. A., Laprie, J.-C., Randell, B., & Kaunas, V. M. U. (2001). *Fundamental Concepts of Dependability* (Department of Computing Science Technical Report Series No. 739). http://www.cs.ncl.ac.uk/publications/trs/papers/739.pdf

Baumeister, R. F., Bratslavsky, E., Finkenauer, C., & Vohs, K. D. (2001). Bad is stronger than good. *Review of General Psychology*, *5*(4), 323–370. https://doi.org/10.1037//1089-2680.5.4.323

Bhatia, A., Chandani, A., & Chhateja, J. (2020). Robo advisory and its potential in addressing the behavioral biases of investors — A qualitative study in Indian context. *Journal of Behavioral and Experimental Finance*, *25*, 100281. https://doi.org/10.1016/j.jbef.2020.100281

Cardillo, G., & Chiappini, H. (2024). Robo-advisors: A systematic literature review. *Finance Research Letters*, *62*, 105119. https://doi.org/10.1016/j.frl.2024.105119

Carroll, J. M., & Olson, J. R. (1988). Mental Models in Human-Computer Interaction. In M. Helander (Ed.), *Handbook of Human-Computer Interaction* (pp. 45–65). North Holland. https://doi.org/10.1016/B978-0-444-70536-5.50007-5

Choudhury, A., & Shamszare, H. (2023). Investigating the Impact of User Trust on the Adoption and Use of ChatGPT: Survey Analysis. *Journal of Medical Internet Research*, *25*, e47184. https://doi.org/10.2196/47184

Colville, S., & Ostern, N. (2024). Trust and Distrust in GAI Applications: The Role of AI Literacy and Metaknowledge. In Association for Information Systems (Chair), *ICIS 2024.* Symposium conducted at the meeting of Association for Information Systems, Bankok, Thailand.

D'Acunto, F., Prabhala, N., & Rossi, A. G. (2019). The Promises and Pitfalls of Robo-Advising. *The Review of Financial Studies*, *32*(5), 1983–2020. https://doi.org/10.1093/rfs/hhz014

Endsley, M. R. (2023). Supporting Human-AI Teams: Transparency, explainability, and situation awareness. *Computers in Human Behavior*, *140*, Article 107574. https://doi.org/10.1016/j.chb.2022.107574

Funder, D. C. (1995). On the accuracy of personality judgment: A realistic approach. *Psychological Review*, *102*(4), 652–670. https://doi.org/10.1037/0033-295x.102.4.652

Hodge, F. D., Mendoza, K. I., & Sinha, R. K. (2021). The Effect of Humanizing Robo-Advisors on Investor Judgments *Contemporary Accounting Research*, *38*(1), 770–792. https://doi.org/10.1111/1911-3846.12641

Hoff, K. A., & Bashir, M. (2015). Trust in automation: Integrating empirical evidence on factors that influence trust. *Human Factors*, *57*(3), 407–434. https://doi.org/10.1177/0018720814547570

426

Hohenberger, C., Lee, C., & Coughlin, J. F. (2019). Acceptance of robo-advisors: Effects of financial experience, affective reactions, and self-enhancement motives. *Financial Planning Review*, *2*(2), Article e1047. https://doi.org/10.1002/cfp2.1047

Houston, R. W., & Taylor, G. K. (1999). Consumer Perceptions of CPA WebTrust SM Assurances: Evidence of an Expectation Gap. *International Journal of Auditing*, *3*(2), 89–105. https://doi.org/10.1111/1099-1123.00051

Isaia, E., & Oggero, N. (2022). The potential use of robo-advisors among the young generation: Evidence from Italy. *Finance Research Letters*, *48*, 103046. https://doi.org/10.1016/j.frl.2022.103046

Johnson-Laird, P. N. (2010). Mental models and human reasoning. *Proceedings of the National Academy of Sciences of the United States of America*, *107*(43), 18243–18250. https://doi.org/10.1073/pnas.1012933107

Johnson-Laird, P. N. (1980). Mental Models in Cognitive Science. *Cognitive Science*, *4*(1), 71–115. https://doi.org/10.1207/s15516709cog0401_4

Kelly, S., Kaye, S.-A., & Oviedo-Trespalacios, O. (2023). What factors contribute to the acceptance of artificial intelligence? A systematic review. *Telematics and Informatics*, *77*, 101925. https://doi.org/10.1016/j.tele.2022.101925

Kim, D. J., Ferrin, D. L., & Rao, H. R. (2008). A trust-based consumer decision-making model in electronic commerce: The role of trust, perceived risk, and their antecedents. *Decision Support Systems*, *44*(2), 544–564. https://doi.org/10.1016/j.dss.2007.07.001

Kim, D. J., Yim, M.-S., Sugumaran, V., & Rao, H. R. (2016). Web assurance seal services, trust and consumers' concerns: an investigation of e-commerce transaction intentions across two nations. *European Journal of Information Systems*, *25*(3), 252–273. https://doi.org/10.1057/ejis.2015.16

Kirlappos, I., Sasse, M. A., & Harvey, N. (2012). Why Trust Seals Don't Work: A Study of User Perceptions and Behavior. In S. Katzenbeisser (Ed.), *Lecture Notes in Computer Science: Vol. 7344. Trust and trustworthy computing: 5th international conference, TRUST 2012, Vienna, Austria, June 13 - 15, 2012 ; proceedings* (Vol. 7344, pp. 308–324). Springer. https://doi.org/10.1007/978-3-642-30921-2_18

Kirmani, A., & Rao, A. R. (2000). No Pain, No Gain: A Critical Review of the Literature on Signaling Unobservable Product Quality. *Journal of Marketing*, *64*(2), 66–79. https://doi.org/10.1509/jmkg.64.2.66.18000

Laugwitz, B., Schrepp, M., & Held, T. (2006). Konstruktion eines Fragebogens zur Messung der User Experience von Softwareprodukten. In H. M. Heinecke, H. Paul, A. M. Heinecke, & H. Paul (Eds.), *Mensch & Computer 2006: Mensch und Computer im StrukturWandel ; [Tagung "Mensch & Computer" 2006 ; Tagungsband* (pp. 125–134). Oldenbourg. https://doi.org/10.1524/9783486841749.125

Lee, J. D., & See, K. A. (2004). Trust in automation: Designing for appropriate reliance. *Human Factors*, *46*(1), 50–80. https://doi.org/10.1518/hfes.46.1.50_30392

Lewicki, R. J., & Wiethoff, C. (2000). Trust, trust development, and trust repair. In M. Deutsch & P. T. Coleman (Eds.), *The handbook of conflict resolution: Theory and practice* (pp. 86–107). Jossey-Bass/Wiley.

Liao, Q., & Sundar, S. S. (2022). Designing for Responsible Trust in AI Systems: A Communication Perspective. In *FAccT '22: Proceedings of the 2022 ACM Conference on Fairness, Accountability, and Transparency* (pp. 1257–1268). ACM. https://doi.org/10.1145/3531146.3533182

Lins, S [Sebastian], Greulich, M., Löbbers, J., Benlian, A [Alexander], & Sunyaev, A [Ali] (2024). Why so skeptical? Investigating the emergence and consequences of consumer skepticism toward web seals. *Information & Management*, *61*(2), 103920. https://doi.org/10.1016/j.im.2024.103920

Lowry, P. B., Moody, G., Vance, A., Jensen, M., Jenkins, J., & Wells, T. (2012). Using an elaboration likelihood approach to better understand the persuasiveness of website privacy assurance cues for online consumers. *Journal of the American Society for Information Science and Technology*, *63*(4), 755–776. https://doi.org/10.1002/asi.21705

Lowry, P. B., Zhang, D., & Wu, D. (2014). Understanding patients' compliance behavior in a mobile healthcare system: The role of trust and planned behavior. *International Conference on Information Systems (ICIS 2014) Auckland, New Zealand*. https://papers.ssrn.com/sol3/papers.cfm?abstract_id=2529599

Madhavan, P., & Wiegmann, D. A. (2007). Similarities and differences between human–human and human–automation trust: an integrative review. *Theoretical Issues in Ergonomics Science*, *8*(4), 277–301. https://doi.org/10.1080/14639220500337708

Morris, N. M., & Rouse, W. B. (1988, January 8). *Human operator response to error-likely situations in complex engineering systems* (T-3227). NASA. https://ntrs.nasa.gov/citations/19890008637

Parasuraman, R., & Riley, V. (1997). Humans and Automation: Use, Misuse, Disuse, Abuse. *Human Factors*(39(2)), 230–253.

Petty, R. E., & Cacioppo, J. T. (1984). Source factors and the elaboration likelihood model of persuasion. *Advances in Consumer Research*, *11*, 668–672. http://www.communicationcache.com/uploads/1/0/8/8/10887248/source_factors_and_the_elaboration_likelihood_model_of_persuasion.pdf

PricewaterhouseCoopers. (May 2020). *How mature is AI adoption in financial services? A PwC Study across the DACH region*. https://www.pwc.de/de/future-of-finance/how-mature-is-ai-adoption-in-financial-services.pdf

Saeed, W., & Omlin, C. (2023). Explainable AI (XAI): A systematic meta-survey of current challenges and future opportunities. *Knowledge-Based Systems*, *263*, 110273. https://doi.org/10.1016/j.knosys.2023.110273

Schlicker, N., Baum, K., Uhde, A., Sterz, S., Hirsch, M. C., & Langer, M. (2025). How do we assess the trustworthiness of AI? Introducing the trustworthiness assessment model (TrAM). *Computers in Human Behavior*, *170*, Article 108671. https://doi.org/10.1016/j.chb.2025.108671

Schrepp, M. (September 2023). *User Experience Questionnaire Handbook*. https://www.ueq-online.org/Material/Handbook.pdf

Stibe, A., Krüger, N., & Behne, A. (2022). Knowledge Behavior Gap Model: An Application for Technology Acceptance. In I. Awan, M. Younas, & A. Poniszewska-Marańda (Eds.), *Lecture Notes in Computer Science. Mobile Web and Intelligent Information Systems* (Vol. 13475, pp. 3–17). Springer International Publishing. https://doi.org/10.1007/978-3-031-14391-5_1

Tao, R., Su, C.-W., Xiao, Y., Dai, K., & Khalid, F. (2021). Robo advisors, algorithmic trading and investment management: Wonders of fourth industrial revolution in financial markets. *Technological Forecasting and Social Change*, *163*, 120421. https://doi.org/10.1016/j.techfore.2020.120421

Tausch, A. (2025). *Mental Model of Collaborative Technologies Scale (MMCT-S)*. https://doi.org/10.17605/OSF.IO/HQYN4

Todd, T. M., & Seay, M. C. (2020). Financial attributes, financial behaviors, financial-advisor-use beliefs, and investing characteristics associated with having used a robo-advisor. *Financial Planning Review*, *3*(3), Article e1104. https://doi.org/10.1002/cfp2.1104

Turow, J. (2023). *Americans and Online Privacy: The System is Broken* (Departmental Papers (ASC)). Annenberg Public Policy Center of the University of Pennsylvania. https://repository.upenn.edu/handle/20.500.14332/2058

Visser, E. de, Cohen, M., Freedy, A., & Parasuraman, R. (2014). A Design Methodology for Trust Cue Calibration in Cognitive Agents. In R. Shumaker & S. Lackey (Eds.), *Lecture Notes in Computer Science: Vol. 8526, Virtual, Augmented and Mixed Reality. Applications of Virtual and Augmented Reality* (pp. 251–262). Springer.

Westbrook, L. (2006). Mental models: a theoretical overview and preliminary study. *Journal of Information Science*, *32*(6), 563–579. https://doi.org/10.1177/0165551506068134

Wischnewski, M., Doebler, P., & Krämer, N. (2025). Development and validation of the Trust in AI Scale (TAIS). *OSF*. Advance online publication. https://doi.org/10.31234/osf.io/eqa9y_v1

Wischnewski, M., Krämer, N., Janiesch, C., Müller, E., Schnitzler, T., & Newen, C. (2024). In Seal We Trust? Investigating the Effect of Certifications on Perceived Trustworthiness of AI Systems. *Human-Machine Communication*, *8*, 141–162. https://doi.org/10.30658/hmc.8.7

Wischnewski, M., Krämer, N., & Müller, E. (2023). Measuring and understanding trust calibrations for automated systems: A survey of the state-of-the-art and future directions. In A. Schmidt (Ed.), *ACM Digital Library, Proceedings of the 2023 CHI Conference on Human Factors in Computing Systems* (pp. 1–16). Association for Computing Machinery.

Yi, T. Z., Rom, N. A. M., Hassan, N. M., Samsurijan, M. S., & Ebekozien, A. (2023). The Adoption of Robo-Advisory among Millennials in the 21st Century: Trust, Usability and Knowledge Perception. *Sustainability*, *15*(7), 6016. https://doi.org/10.3390/su15076016